\documentclass[runningheads]{llncs}
\usepackage[T1]{fontenc}
\usepackage{graphicx}

\usepackage{graphics,graphicx} 
\usepackage{amsmath,amssymb}
\usepackage{booktabs}
\usepackage{cite}
\usepackage{subcaption}
\usepackage[font=footnotesize]{caption}
\usepackage{romannum}
\usepackage{comment}
\usepackage{algorithm2e}
\usepackage{chemarrow}
\begin{document}
\title{Dynamical Inference of Cell Size Regulation Parameters}
%
%
\author{C\'esar Nieto \inst{1}\orcidID{0000-0001-6504-4619} \and Sayeh Rezaee\inst{1}\orcidID{0000-0003-2826-4756
} \and Cesar Augusto Vargas-Garcia\inst{2}\orcidID{0000-0002-4286-8882} \and Abhyudai Singh\inst{3}\orcidID{0000-0002-1451-2838}}
%
\authorrunning{ Nieto. C\'esar \textit{et. al.}}
%
\institute{Department of Electrical and Computer Engineering, University of Delaware, Newark, DE, USA \and AGROSAVIA - Corporacion Colombiana de Investigacion Agropecuaria, Mosquera, Cundinamarca, Colombia.\\ \and Department of Electrical and Computer Engineering, Biomedical Engineering, Mathematical Sciences, Center of Bioinformatic and Computational Biology, University of Delaware, Newark, DE, USA\\
\email{\{absingh\}@udel.edu}}
\maketitle          
\begin{abstract}

Cells achieve size homeostasis by regulating their division timing based on their size, added size, and cell cycle time.~Previous research under steady-state conditions demonstrated the robustness of these mechanisms.~However, their dynamic responses in fluctuating environments, such as nutrient depletion due to population growth, remain challenging to fully characterize.~Currently, advances in single-cell microscopy have revealed various cellular division strategies whose underlying molecular mechanisms are complex and not always available.~This study introduces a novel approach to model cell size dynamics using a piecewise deterministic Markov chain framework, where cell division events are modeled as stochastic jumps determined by a division propensity dependent on both current cell size and added size since birth.~We propose a three-parameter characterization for the division process: scale (target added size at division), shape (division stochasticity), and division strategy (relevance of cell size, added size, or cell cycle duration).~We derive analytical formulas for the probability of division, and with this probability, we develop a maximum likelihood estimation (MLE) framework.~We implement a systematic investigation of the accuracy of inference as a function of sample size.~The model's performance is studied across various scenarios, including those exhibiting dynamical changes in one or more parameters, suggesting its broad applicability for analyzing new experimental data on cell size regulation in dynamic environments.

\keywords{Systems biology  \and Maximum likelihood estimation \and Stochastic hybrid systems.}
\end{abstract}

Cell size regulation is one of the processes that microbes have optimized to survive in their environment~\cite{jun2018fundamental}.~This regulation is the result of two processes: Cell growth and cell division~\cite{lin2018homeostasis}.~Cell growth is related to metabolic performance including fundamental internal processes such as nutrient assimilation, and ion concentration maintenance, whereas cell division defines the cell proliferation rate~\cite{amir2014cell,si2019mechanistic}.~Consequently, cell size dynamics provide a simple yet powerful framework for exploring how effectively microbes grow and reproduce under varying environmental conditions, including nutrient depletion, pH changes, osmotic shocks, and exposure to antibiotics \cite{nguyen2021distinct}. Moreover, fluctuations in cell size regulation contribute to noise in gene expression, potentially impairing cellular functionality and influencing population fitness by altering rates of cell elongation and DNA replication~\cite{soltani2016intercellular,sanchez2013regulation}.

The investigation of cell size regulation under variable nutrient conditions is particularly relevant for microbiology and biotechnology.~Natural growth environments are typically dynamic, presenting a continuum of nutrient availability that requires real-time cellular size adjustments.~Advances in single-cell trapping, tracking, and high-resolution imaging techniques have elucidated significant changes in bacterial cell volume in different and dynamical environments~\cite{shi2021precise}.~For example, when bacteria grow in rich media, they have a relatively large cell, while they become smaller under nutrient-limited conditions~\cite{bakshi2021tracking}.~In such fluctuating environments, cell volume is determined by a trade-off between resource allocation for biomass production and the timely initiation of cell division~\cite{cylke2024energy,kratz2023dynamic}.~Understanding the dynamics and molecular mechanisms of this modulation is important for approaching the synchronization of processes related to growth and division, including cell elongation, chromosome replication, and protein synthesis, which collectively facilitate rapid adaptation to new nutrient environments~\cite{nguyen2021distinct}.~
When the complete mechanism of cell division is unknown, a mathematical framework is necessary that enables comparison of the observation with cases where the mechanisms are well-characterized through these coarse-grained models~\cite{nieto2025generalized}.

The mechanisms coupling cell growth with division timing are commonly categorized into three archetypal strategies: the "sizer" where cell division occurs upon reaching a critical, predetermined size; the "adder" where cells add a consistent amount of size during each cell cycle, independent of their initial size; and the "timer" where cell division is initiated after a fixed duration, regardless of the cellular dimensions~\cite{miotto2024size,vargas2020modeling,jones2023first}.~These strategies are empirically discriminated by the linear relationship between the cell size added during a cell cycle (added size at division) and the cell size at the beginning of the cycle (size at birth).~A slope of -1 characterizes a sizer, a slope of 0 describes the adder, and a slope of 1 indicates the timer.~Although these ideal strategies offer a valuable descriptive framework, the reality of biological systems often reveals more complex, intermediate behaviors, such as "sizer-adder" (slopes between -1 and 0) and "timer-adder" (slopes between~0~and~1).~Some organisms such as certain yeasts, as well as slow-growing \textit{E.~coli} and \textit{Mycobacteria}, have been suggested to exhibit sizer-adder characteristics, whereas other bacteria such as \textit{C.~crescentus} and \textit{C.~difficile} demonstrate timer-adder division\cite{liu2023cell,miller2023fission,ribis2024unique}.~Advances in experimental techniques, particularly those that allow mass measurement in complex human cell lines, have revealed a wide range of growth and division strategies between species~\cite{cadart2018size}.

While effective in describing experimental observations, these division strategies often pose challenges for direct extension into a continuous-time mathematical framework.~Specifically, they do not inherently allow the estimation of the cell's probability of division at any given size and time since birth.~To overcome this limitation, a more robust approach adopts the stochastic hybrid systems (SHS) framework~\cite{singh2010stochastic}.~In SHS, cell size is modeled as a continuously growing variable, while division events are represented as discrete "jumps" (typically halving the cell size) that occur at a specific rate, known as the division propensity.~This propensity captures essential information about how the division process evolves over time and can be conceptualized as an effective process linked to the progression through various cell cycle stages or the accumulation of key division-regulatory molecules~\cite{jia2021cell,nieto2021continuous}.~However, acquiring precise information on internal cell cycle stages or the concentration of specific division regulators is often experimentally challenging~\cite{panlilio2021threshold,si2019mechanistic}.~Therefore, an SHS approach that only considers readily observable variables such as cell size, added size, and cell cycle time is more practical.~Additionally, by defining both the division propensity and the growth rate, the continuous-time dynamics of the cell size distribution can be rigorously estimated by solving the associated Chapman-Kolmogorov equation (CKE)~\cite{nieto2021continuous,nieto2025joint}.~Recent contributions reveal that the cell size dynamics predicted by these division propensities matches with the experiments~\cite{rezaee2025inferring, nieto2025joint}.~Although parameter inference has been performed under steady conditions for different cells~\cite{jia2021cell}, the development of an algorithm for inferring the division parameters in any arbitrary condition out of steady state is still an open problem.

In this contribution, we study a recently developed family of division propensities based on cell size and added size since birth \cite{nieto2025generalized}.~We demonstrate how these proposed functions directly relate to the classical division strategies and develop inference methods to determine their parameters using observed statistics of cell size, added size, growth rate, and occurrence of division within an infinitesimal time interval.~We derive explicit formulas for the log-likelihood functions and study their performance in simulated datasets with known parameters, and study their performance in different time-dependent scenarios.~Through this solution, our objective is to establish a more comprehensive and quantitative framework to understand the regulation of bacterial cell size.
 
\begin{figure*}[!ht]
\centering
\includegraphics[width=1\linewidth]{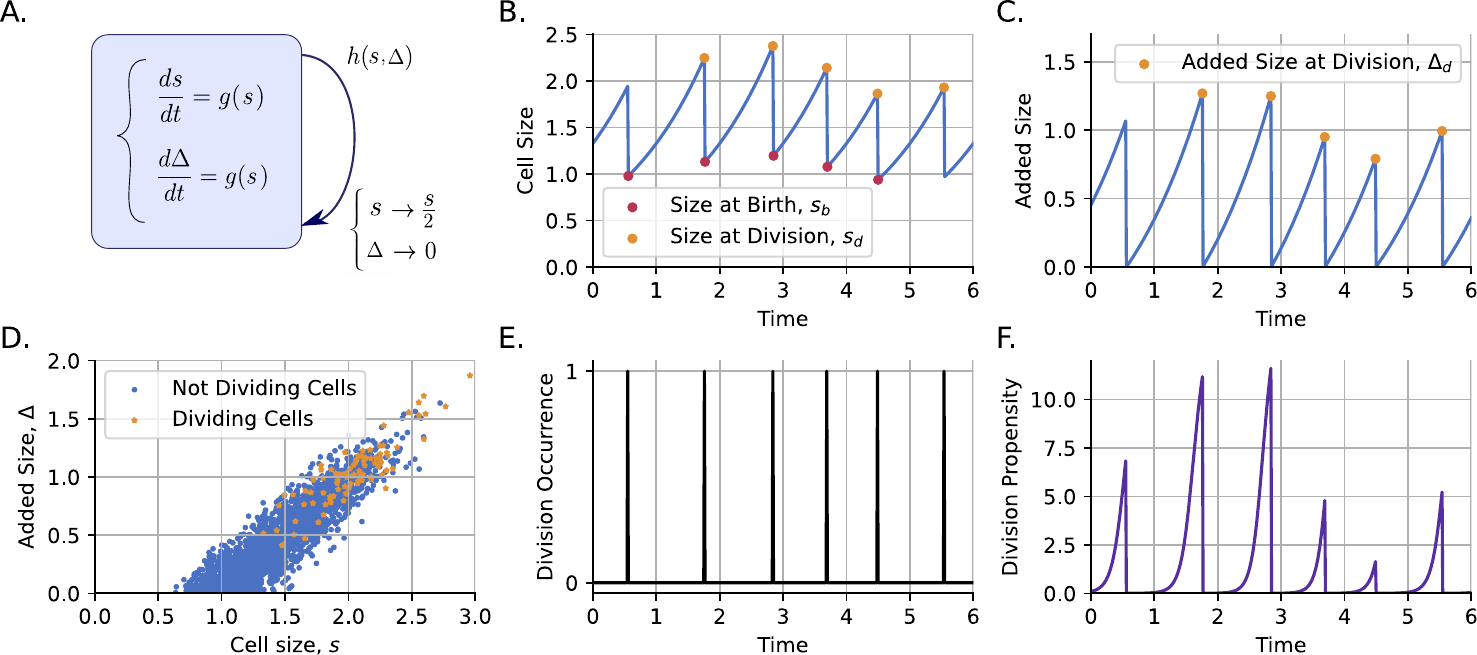}
\caption{\small \textbf{Cell size dynamics and the visualization of the main variables defining the cell size control.} (\textbf{A})~Diagram explaining the cell division process as a stochastic hybrid system.~(\textbf{B})~Simulated example of a cell size trajectory consisting of exponential growth with discrete jumps each halving the cell size representing the cell division.~The cell cycle is defined as the processes occurring in between two consecutive divisions.~The size at the beginning of cell cycle (red dots) is defined as the size at birth $s_b$, the cell size at the end of the cell cycle (orange dots) is cell size at division $s_d$.~(\textbf{C})~Added size $\Delta$ as a function of time (solid line) and the added size at division $\Delta_d$ (orange dots).~(\textbf{D})~Example of a dataset used for the inference.~Each point corresponds to a cell with given size $s$ and added size $\Delta$ with two possible states: Either the cell is not dividing (blue dots) or the cell is dividing (Orange squares).~(\textbf{E})~Example of the division occurrence function.~This function takes the value of 0 if the cell does not divide during the time interval $(t_0,t_0+\delta t])$ and the value of 1 if division occurs during that interval. In this simulation, $\delta t=0.01$.~(\textbf{F})~Division propensity as a function of time for the example cell illustrated in panels~(B)~and (C).}
\label{fig:fig1}
\end{figure*}

\section{Cell Size Dynamics: Solution for Adder Division Strategy}

In this section, we introduce the main concepts of our cell size dynamics model, including the primary variables, their evolution, and the characteristics of division propensity in the simplest division strategy: adder.~In our approach, cell size dynamics is described by a continuous exponential growth with discrete jumps (halving the cell size) representing the division.~We define two main variables:
\begin{itemize}
\item \textit{Cell size}, $s>0$:
This random process describes the continuous growth of a cell over time. At the start of each cycle, the cell has a size 
$s_b$ (size at birth). It then grows continuously until it reaches the size at division 
$s_d$, where division occurs. At division, the cell splits into two daughter cells, each inheriting half of the parent’s size. In our analysis, we follow a single lineage by tracking one of the daughter cells through successive cycles.
\item \textit{Added size}, $\Delta\geq~0$:
This random process quantifies the increase in cell size from birth to division.~Starting at zero when the cell is born, $\Delta$ grows at the same rate as $s$, and reaches $\Delta_d$, (added size at division) just before division.~After division, $\Delta$ resets to zero, 
marking the start of a new cycle.~By definition, $\Delta = s - s_b$, and thus, $\Delta_d = s_d - s_b$, with $s_b$ varying from cycle~to~cycle.  
\end{itemize}

The relationship between random variables $s_b$, $s_d$, and $\Delta_d$ defines the division strategy.~Traditionally, the mean added  size at division given the size at birth~$s_b$, denoted by $\langle\Delta_d|s_b\rangle$, is  phenomenologically approximated as a linear function of~$s_b$:
\begin{equation}\label{eq:meansizer}
    \langle\Delta_d|s_b\rangle=\alpha(s_b-\langle s_b\rangle)+\langle \Delta_d\rangle,\
\end{equation}
in which the slope $\alpha$ characterizes the division strategy.~For the adder, $\alpha=0$; for the timer-adder, $0<\alpha<1$; and, finally, for the sizer-adder, $-1<\alpha<0$. Additionally, $\langle s_b\rangle$ is the mean size at birth, and $\langle \Delta_d\rangle$ is the mean added size at division.~Next, we will explore how to connect our continuous-time approach with this discrete mapping framework.

The dynamics of cell size variables are described by the following system of differential equations:
\begin{equation}\label{eq:system}
\dfrac{ds}{dt}=g( {s}),\quad
\dfrac{d\Delta}{dt}=g( {s}).
\end{equation}
Here, $g(s)>0$ represents the \textit{growth law}~\cite{nieto2025generalized}, an arbitrary continuous function of $s$ that describes the rate of cell growth.~A simple particular case of a growth law is exponential growth $g(s)=\mu s$, where $\mu$ denotes the exponential growth rate.~Upon division, the cell variables reset according to the following map.
\begin{equation}\label{eq:maps}
s\rightarrow s/2,\quad \Delta \rightarrow 0.
\end{equation}

Cell division is modeled as a random jump process, where the probability of division occurring in an infinitesimal interval $(t,t+dt]$ is given by $h(s,\Delta)dt$, with
$h(s,\Delta)$ representing the \textit{division rate} or \textit{division propensity}.~
In~\cite{nieto2025generalized}, authors showed that the adder mechanism emerges when the division rate takes the form:
\begin{equation} \label{eq:division}
h(s,\Delta) = g( {s}) \lambda( {s,\Delta}) ,
\end{equation}
where $\lambda(s,\Delta)>0$ denotes the \textit{division function}.~For the adder mechanism, division function $\lambda$ only depends on $\Delta$, while in the general case, $\lambda$ can be a function of both $s$ and $\Delta$.~

\noindent

After defining the division propensity, we now describe the probability that a division occurs over an arbitrarily finite time interval $(t_0,t_0+\delta t)$, where $t_0$ is an arbitrary time instant and $\delta t>0$ is a small but not infinitesimal time interval, usually the sampling period. This interval 
$\delta t$ is assumed to be small enough that more than one division in the same interval is negligibly unlikely; therefore, each interval contains either zero or one division event.

Given a cell with size $s(t_0)$ and added size $\Delta(t_0)$ at the beginning of the interval, the division probability is:
\begin{equation}\label{eq:pdiv}
    \mathbb{P}(\text{Division occurs in }(t_0,t_0+\delta t)|{s(t_0),\Delta(t_0)})= 1-e^{-\int^{t_0+\delta t}_{t_0} h(s(t'),\Delta(t'))dt'}.
\end{equation}
This expression follows from the master equation of division events as a time-inhomogeneous Poisson process whose rate $h(s,\Delta)$ varies according to the evolving cell size and added size~\cite{nieto2025generalized}. To evaluate the integral in \eqref{eq:pdiv}, we solve the dynamical equations for 
$s(t)$ and $\Delta(t)$ introduced in \eqref{eq:system}, and therefore $h(s,\Delta)$ along the integration interval. In the simplest approximation, the growth rate is fitted to exponential $\mu$ during $(t_0,t_0+\delta t)$ and therefore, the only required inputs to solve \eqref{eq:pdiv} are $s(t_0)$ and $\Delta(t_0)$. 

To compare the proposed framework with experimental results, a modification is necessary. Since experimental data are recorded at discrete time points rather than continuously, we must adapt the continuous formulation to match this discrete-time structure. We will shortly explain this modification next.

\subsection{Implementing the framework in a discrete-time setting}

Experimental trajectories consist of measurements collected at discrete time~points $(t_1,t_2,\cdots,t_n,\cdots, t_N)$ 
with a fixed acquisition interval $\delta t=t_{n+1}-t_n$~and $N$ is the total number of data points. As a result, the observed cell size trajectories are $(s(t_1),s(t_2),\cdots,s(t_n),\cdots, s(t_N))$ and added size are $(\Delta(t_1),\Delta(t_2),\cdots,\Delta(t_n),\cdots, \Delta(t_N))$.

As we assumed each interval $(t_n,t_{n+1})$
 can capture at most one division event, the presence or absence of division within each interval can therefore be represented directly from the data as a binary indicator, taking the value 1 when division occurs between 
$t_n$ and $t_{n+1}$
 and 0 otherwise (See Fig. \ref{fig:fig1}E).
To compute the associated probability of observing a division in a given interval, we apply the continuous-time expression \eqref{eq:pdiv} to the interval $(t_n,t_{n+1})$:
\begin{equation}\label{eq:pdiv1}
    \mathbb{P}(\text{Division occurs in }(t_n,t_{n+1})|{s(t_n),\Delta(t_n)})= 1-e^{-\int^{t_{n+1}}_{t_{n}} h(s(t'),\Delta(t')) dt'}.
\end{equation}

Evaluating this integral requires specifying how 
$\Delta$ evolves between two measurement points. The growth rate 
$\mu$ can be experimentally estimated
by fitting cell size trajectories or measuring biomass growth; thus, it is known at each time $t_n$. With this fit, $\mu$ is assumed to remain constant between successive observations so that the size dynamics can be interpolated smoothly within each interval. This approximation enables a continuous evaluation of the integral even though the data themselves are discrete. In the next section, we show how this probability can be computed explicitly when the division rate $h(s,\Delta)$ is given by a specific analytical expression, which enables the construction of a maximum likelihood estimator for the division parameters.

\section{Characterizing Division Process: The Sigmoidal Division Function}

The functional form of the division rate $h$ is determined by the underlying metabolic mechanism~\cite{nieto2025generalized} or by phenomenological fits to experimental data~\cite{xia2020pde}.~In our theory, there is no reason to choose any particular function since we assume that the information about the division process (the state of the division molecular mechanism) is hidden.~To simplify the calculations and illustrate the concept, we propose a basic example for the division function $\lambda$ following a numerically stable \textit{sigmoidal-shaped} division function:
\begin{equation}\label{eq:sigmo}
    \lambda( {\Delta}) = \dfrac{k}{1+e^{-k( {\Delta}-\Delta_0)}},
\end{equation}
valid for $\Delta>0$ and with $k$ being the \textit{shape parameter} and $\Delta_0$ the \textit{scale~parameter}.~Using the method explained through expression~\eqref{eq:pdiv} and assuming exponential growth with rate $\mu$, we calculate the probability of division given a cell size $s$ and added size $\Delta$ during the finite time interval~$(t_0,t_0+\delta t)$:
\begin{equation}\label{eq:prob}
    \mathbb{P}(\text{Division occurs in }(t_0,t_0+\delta t)|{s(t_0),\Delta(t_0)})=
 1 - \frac{1 + e^{-k(\Delta - \Delta_0)}}{e^{-k(\Delta - \Delta_0)} + e^{-ks(1 - e^{\mu \delta t})}}.~\end{equation}

\subsection{Interpretation of division function parameters with respect to measurable statistics}

To represent the model parameters $k$ and $\Delta_0$, as a function of measurable statistics, we can calculate the statistics of added size at division $ {\Delta_d}$. With a similar approach to~\cite{nieto2025generalized}, we can estimate the cumulative distribution function (CDF) of added size at division $ {\Delta_d}$:
\begin{equation} \label{eq:cdfDel}
F_{\Delta_d}(y) :=\mathbb{P}( {\Delta_d}<y | {s_b})= 1-e^{-\int^y_0 \lambda(z) dz}.
\end{equation}
Next, using the CDF, we can find the corresponding probability density function (PDF) of added size at division:
\begin{equation} \label{eq:pdfDel}
f_{ {\Delta_d}}(y) :=\dfrac{d F_{ {\Delta_d}}(y)}{d {\Delta_d}}= \lambda(y) e^{-\int^y_0 \lambda(z) dz}.
\end{equation}
The definition of this PDF is used to obtain the statistical moments of $\Delta_d$: the mean $\langle  {\Delta}_d \rangle$; and the squared coefficient of variation $CV_{ {\Delta_d}}^2$, which quantifies randomness:
\begin{equation}\label{eq:stats0}
\langle \Delta_d \rangle := \int^\infty_0 \Delta_d f_{ {\Delta_d}}(y)dy \quad , \quad
CV_{ {\Delta_d}}^2 := \dfrac{\sigma^2_{\Delta_d}}{\langle \Delta_d\rangle ^2},
\end{equation}
where ${\sigma^2_{\Delta_d}={\langle  {\Delta}^ 2_d \rangle-\langle  {\Delta}_d \rangle^2}}$ is the variance of $\Delta_d$.

The PDF of added size at division using the sigmoidal division function, proposed in Equation \eqref{eq:sigmo} can be calculated as:
\begin{equation}\label{eq:rho2}
    f_{ {\Delta_d}}(y) = \dfrac{ k e^{-k y} \left( 1+e^{k \Delta_0}\right)}{\left(1+e^{-k(y-\Delta_0)} \right)^2}.
\end{equation}
Additionally, using \eqref{eq:stats0} and \eqref{eq:rho2}, we obtain the first- and second-order moments~of $\Delta_d$:
\begin{subequations}\label{eq:del0sig1}
\begin{align}\label{eq:del0sig}
          \langle  {\Delta_d} \rangle =& 
    \dfrac{1}{k}e^{-k \Delta_0}\left(1+e^{k \Delta_0} \right) \ln[1+e^{k \Delta_0}], \\
      \langle  {\Delta_d}^2 \rangle =&\dfrac{\left(1+e^{-k \Delta_0} \right)}{3 k^2} ( \pi^2 -6\operatorname{Li}_2\left(\frac{1}{1+ e^{k \Delta_0}}\right)\nonumber\\
    &+3 \ln(1+e^{k \Delta_0})(2k\Delta_0-\ln[1 + e^{k \Delta_0 }]).
\end{align}
\end{subequations}
Here, $\operatorname{Li}_2(.)$ is the \textit{polylogarithm} function of order~2, which can be represented in \textit{Wolfram Mathematica} using ${\operatorname{Li}_2(z):=PolyLog[2,z]}$.~These moments can be simplified in the case when $k\Delta_0>5$, leading to the following:

\begin{equation}\label{eq:del0approx}
        \langle  {\Delta_d} \rangle \approx
\Delta_0, \hspace{1cm}
      CV^2_{\Delta_d} \approx\frac{\pi^2}{3(k\Delta_0)^2};\hspace{1cm}  k\Delta_0>5,
\end{equation}
which is equivalently valid for $CV^2_{\Delta_d}<0.12$; a range observed in most biological datasets~\cite{taheri2015cell}.~Therefore, it is worth mentioning that the division propensity can be approximated in terms of measurable statistics from data; mean $\langle \Delta_d\rangle$ and noise in added size at division $CV_{\Delta_d}$, as follows. 
\begin{eqnarray}\label{eq:sigmo2}
    h(\Delta,s) &=& \mu s \lambda(\Delta)\nonumber\\ &\approx &\frac{\pi \mu s}{\sqrt{3}\langle \Delta_d \rangle} \frac{1}{CV_{\Delta_d} \left(1 + \exp\left[{-\frac{\pi}{\sqrt{3}CV_{\Delta_d}}\left(\frac{\Delta}{\langle \Delta_d \rangle}-1\right)}\right]\right)}.
\end{eqnarray}

\begin{figure*}[!ht]
\centering
\includegraphics[width=1\linewidth]{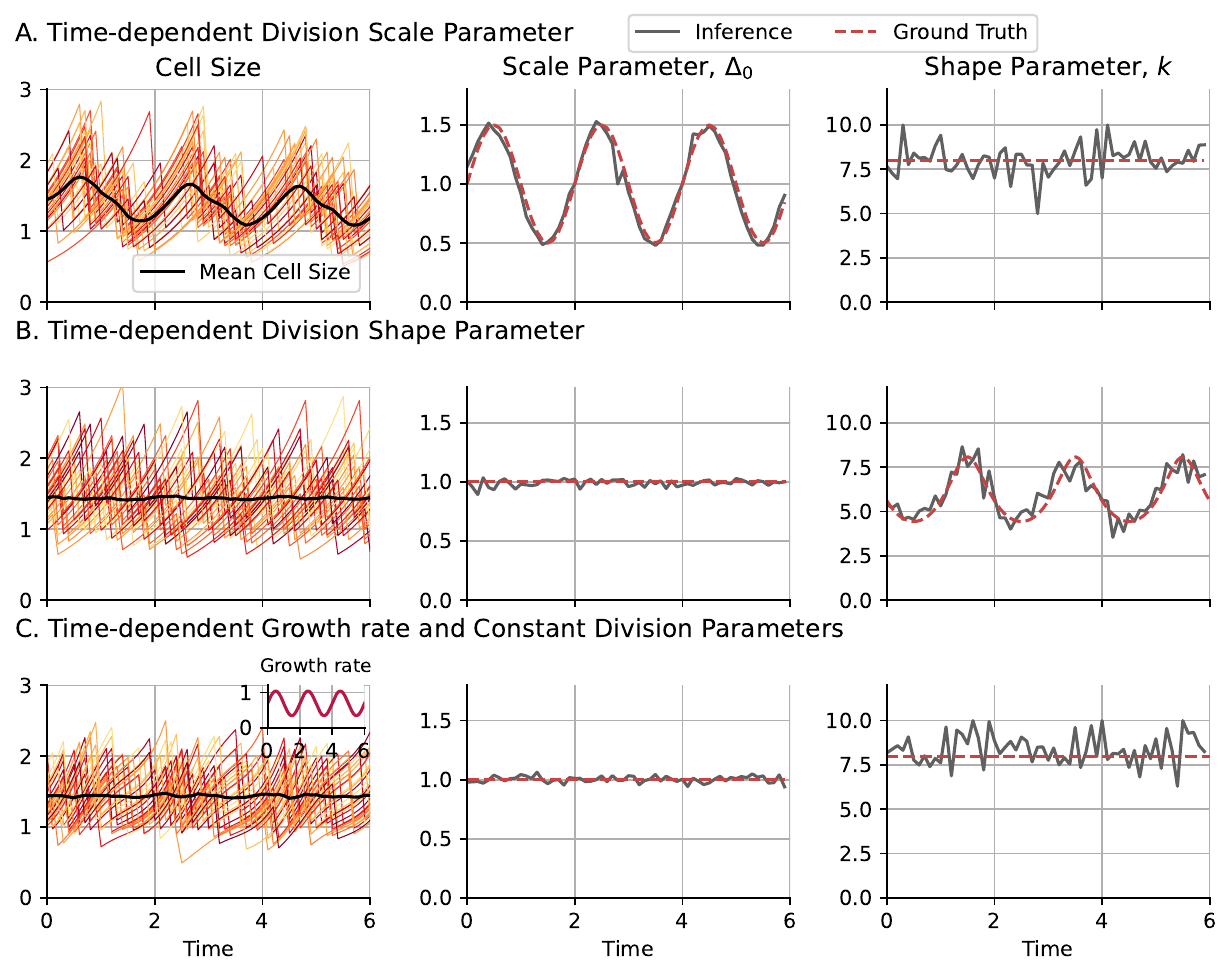}
\caption{{\small \textbf{Inference of division parameters in a dynamical scenario.} We simulate 1000 cells with a dynamical change in different division parameters.~The figure presents (left) example of 30 cell size trajectories (different shades of red), (center) inference of the scale parameter $\Delta_0$ (gray) compared with the ground truth (red dashed line) and (right) results of the inference of the shape parameter $k$ (gray) compared to the ground truth (red dashed line).~(\textbf{A}) Results when the scale parameter is a function of time following $\Delta_0=1+0.5\sin(2\pi t)$, $\mu=\ln(2)$, $k=8$.~(\textbf{B}) Results when the shape parameter is a function of time following $k=6+2\sin(2\pi t)$, $\Delta_0=1$, $\mu=\ln(2)$.~(\textbf{C}) Results when the growth rate is a function of time following $\mu=\ln(2)(1+0.5\sin(2\pi t))$, $\Delta_0=1$, $k=8$.~For all studied cases, $\delta t=0.1$.~ } }
\label{fig:fig2}
\end{figure*}
\subsection{Inference of the Division Parameters}

For a given time $t_n$, experimental data set consists of a set of points ${\mathbf{X}:=\{(\mu(t_n),s(t_n), \Delta(t_n))_i\}}$, where the subindex $i$ represents each cell, and $\mu$, $s$ and $\Delta$ are their respective growth rate, size and added size at time $t_n$. The likelihood function also depends on whether the cell divided or not in the time interval $(t_n,t_{n+1})$ (see Fig.~\ref{fig:fig1}D).
For the adder model, given the model parameters $\theta= (k,\Delta_0)$, we write the log-likelihood function of the division events using equation~\eqref{eq:prob}:
\begin{equation}
   \mathcal{L}_i(\mu,\Delta,s;\theta ):=\left\{
\begin{array}{l}
\ln{\left(1 - \frac{1 + e^{-k(\Delta - \Delta_0)}}{e^{-k(\Delta - \Delta_0)} + e^{-ks(1 - e^{\mu \delta t})}} \right);}\;\;\text{ if division occurs,}\; \;\; \\
\\  \ln{\left(\frac{1 + e^{-k(\Delta - \Delta_0)}}{e^{-k(\Delta - \Delta_0)} + e^{-ks(1 - e^{\mu \delta t})}}\right)\;\;\;\;\;\;;}\; \;\;\; \text{otherwise},
\end{array}
\right.
\end{equation}
which depends parametrically on $\theta$. The maximum likelihood estimation (MLE) method consists of finding $\theta^*=(k^*,\Delta_0^*)$ as the optimal parameters that maximize the total log-likelihood function, estimated using experimental data~$\mathbf{X}$ as follows.
\begin{equation}
    \theta ^ {*} = \operatorname*{argmax}_\theta \left(\sum_{i}\mathcal{L}_i(\mu,\Delta,s ;\theta) \right).
\end{equation}

Notice that the division log-likelihood depends only on the current values of $s$, $\Delta$ and $\mu$ at a given time and does not depend on other variables such as the size at birth $s_b$ or the value of these variables in any previous time.
Consequently, our inference method can estimate the division parameters at any time $t_n$ under the assumption that, although these parameters may vary over time in different scenarios, all cells observed at the same time point $t_n$ share the same division parameter values.


\subsection{Performance of the inference method over a dataset with dynamical division parameters}

To evaluate the robustness of the maximum likelihood inference method under non-stationary conditions, we generated synthetic datasets in which key division parameters vary dynamically over time.~Specifically, we generated trajectories for 1000 cells, each with cell size dynamics governed by a stochastic model incorporating time-dependent division parameters.~We then apply our inference framework to estimate the instantaneous values of the scale $\Delta_0$ and shape $k$ parameters from these simulated trajectories.

Fig.~\ref{fig:fig2} summarizes the inference results across three different scenarios.~In the first case (Fig.~\ref{fig:fig2}A), the scale parameter $\Delta_0$ is varied as a sinusoidal function of time: 
\begin{equation}
    \Delta_0=1+0.5\sin(2\pi t),
\end{equation}
with a fixed shape parameter and growth rate.~The center panel in Fig.~\ref{fig:fig2}.A shows that the inferred $\Delta_0$ closely follows the true underlying function (red dashed line), demonstrating that the inference method can track time-varying division thresholds with high fidelity.
Next, in the second scenario, the shape parameter $k$ is assumed to vary sinusoidally as:
\begin{equation}
k=6+2\sin(2\pi t),
\end{equation}
where the scale parameter and growth rate are constant (Fig.~\ref{fig:fig2}.B).~Again, the inference method successfully recovered the temporal profile of 
$k$, as shown in the right panel.~Finally, we assessed the impact of time-varying growth rates by letting:
\begin{equation}
    \mu=\ln(2)(1+0.5\sin(2\pi t)),
\end{equation}
while keeping $\Delta_0$ and $k$ constant (see Fig.~\ref{fig:fig2}.C).~In this case, although the division parameters remain fixed, changes in the growth rate affect the observed cell size trajectories, which could, in principle, confound inference.~Nonetheless, the inference of both parameters remained stable over time, indicating that the method distinguishes between effects due to intrinsic division variability and extrinsic growth modulation.

In all three scenarios, the sampling resolution was fixed at 
$\delta t=0.1$, ensuring that the inference was performed on realistically sparse time-series data.~These results demonstrate that our inference framework remains accurate and stable even under time-varying conditions, making it suitable for application to biological systems undergoing environmental or regulatory fluctuations.


 \section{Cell Size Control Beyond the Adder}

 \begin{figure*}[!ht]
\centering
\includegraphics[width=1\linewidth]{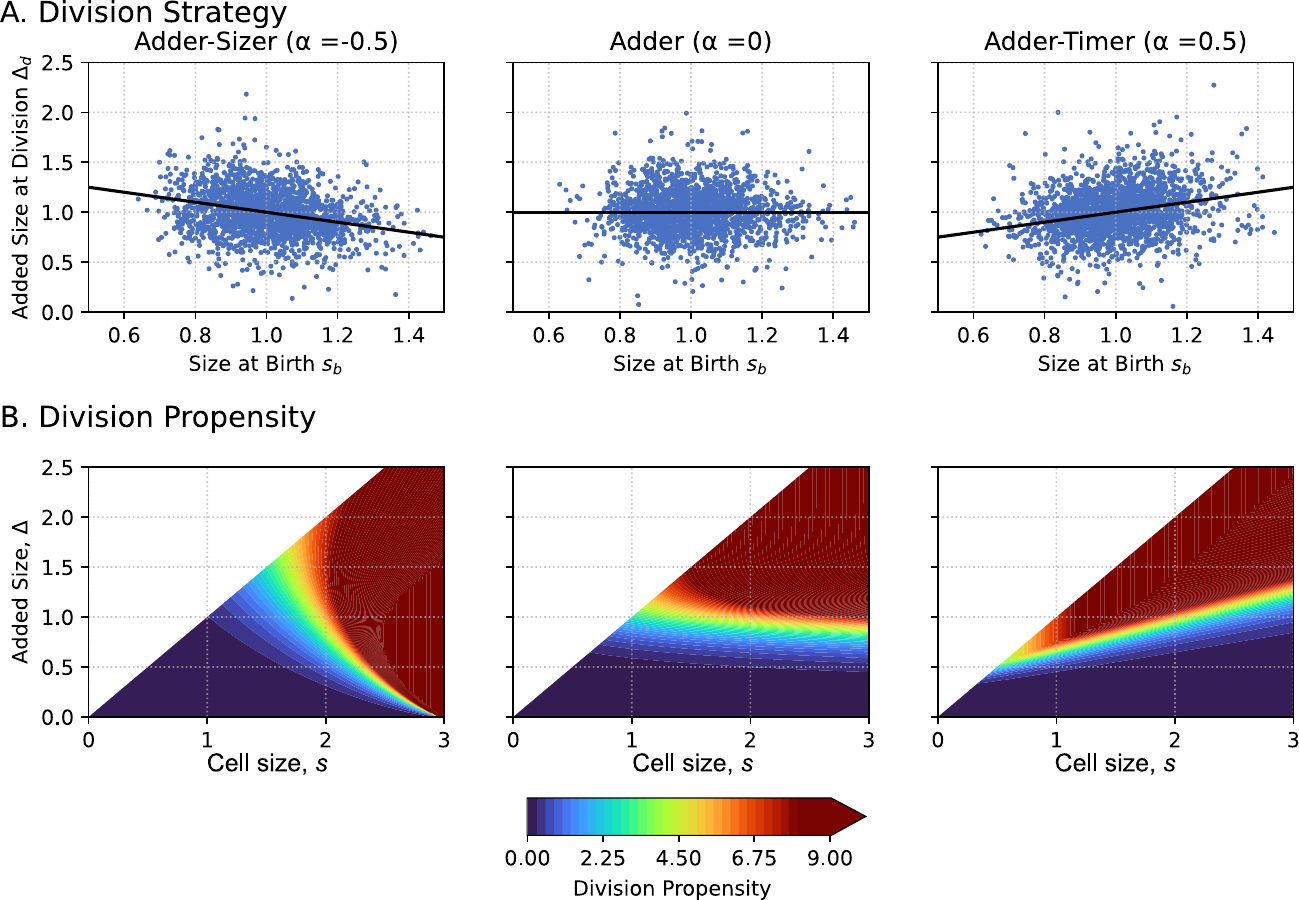}
\caption{{\small \textbf{Different division strategies and the visualization of their respective division propensity as a function of cell size and added size.} (\textbf{A})~Simulated trend of added size at division versus size at birth for different division strategies (different values of $\alpha$).~Black line represents the trend line.~(\textbf{B}) Division propensity as function of $s$ and $\Delta$ (color plot).~Three different strategies are contrasted: sizer-adder ($\alpha=0.5$, left), adder ($\alpha=0$, center) and timer-adder ($\alpha=-0.5$, right).~The division propensity is fitted to the sigmoidal division function with parameters: $\langle\Delta_d\rangle=\langle s_b\rangle=1$, $CV^2_{\Delta_d|s_b}=0.05$, the given $\alpha$ and $\mu=\ln(2)$.} }
\label{fig:fig3}
\end{figure*}

In the previous sections, we proposed division functions and described cell-size statistics, focusing on adder mechanism due to simplicity.~However, a similar approach can be extended to describe the other general division strategies.~In this section, we extend the model to include timer-adder and sizer-adder mechanisms.~In these cases, the mean added size at division $\langle\Delta_d\rangle$ depends on the size at birth $s_b$.~Phenomenologically, it is possible to extend the expression~\eqref{eq:meansizer} including the division randomness to approximate the conditional statistics of $\Delta_d$ given $s_b$ as~\cite{nieto2024mechanisms, nieto2020unification}:
\begin{subequations}\label{eq:mdelta}
    \begin{align}
        \langle \Delta_d|s_b \rangle &= \alpha \left( s_b - \langle s_b \rangle \right) + \langle \Delta_d \rangle,\label{eq:mdelta1}\\
        CV^2 _{\Delta_d|s_b} &= \dfrac{{\sigma^2_{\Delta_d|s_b}}}{\langle \Delta_d|s_b \rangle^2},
    \end{align}
\end{subequations}
where $\langle \Delta_d|s_b \rangle$  and $CV^2 _{\Delta_d|s_b}$ represent the conditional mean and squared coefficient of variation of added size at division $\Delta_d$ given size at birth $s_b$, respectively.~Moreover, $\langle s_b\rangle$ is the mean size at birth, and $\langle \Delta_d\rangle$ is the mean added size at division.~Although the variance of $\Delta_d$ given $s_b$, denoted as $\sigma^2_{\Delta_d|s_b}$, can be any arbitrary function of $s_b$ fitted from experiments, here, we simplify $CV^2 _{\Delta_d|s_b}$ as a constant function independent of $s_b$.~This is satisfied by keeping the product $k\Delta_0$ constant, which is convenient for an easy solution of the general equation of the system~\eqref{eq:del0sig}.~Furthermore, keeping $k\Delta_0$ constant also allows us to perform the approximation \eqref{eq:del0approx} with more confidence.~In this way, we can approximate the division propensity as follows:
{\small
\begin{align}\label{eq:sigmo2sb}
    &h(\Delta,s|s_b) \approx \nonumber \\ &\frac{\pi \mu s/\sqrt{3}}{\alpha \left( s_b - \langle s_b \rangle \right) + \langle \Delta_d \rangle} \frac{1}{CV_{\Delta_d|s_b} \left(1 + \exp\left[{-\frac{\pi}{\sqrt{3}CV_{\Delta_d|s_b}}\left(\frac{\Delta}{\alpha \left( s_b - \langle s_b \rangle \right) + \langle \Delta_d \rangle}-1\right)}\right]\right)}.
\end{align}}
Performing additional approximations, such as $\langle s_b\rangle=\langle \Delta_d \rangle$, valid for steady-state conditions~\cite{modi2017analysis} and identity $s_b:=s-\Delta$, we can write the division propensity \eqref{eq:sigmo2sb} solely in terms of the variables $s$ and $\Delta$ and the observable statistics as follows:
\begin{eqnarray}\label{eq:sigmo3}
    h(\Delta,s) \approx&& \frac{\pi \mu s}{\sqrt{3}\left(\alpha \left( s-\Delta - \langle \Delta_d \rangle \right) + \langle \Delta_d \rangle\right)} \times\nonumber\\&&\frac{1}{CV_{\Delta_d|s_b} \left(1 + \exp\left[{-\frac{\pi}{\sqrt{3}CV_{\Delta_d|s_b}}\left(\frac{\Delta}{\alpha \left( s-\Delta - \langle \Delta_d \rangle \right) + \langle \Delta_d \rangle}-1\right)}\right]\right)}.
\end{eqnarray}
A visualization of the division propensity for different values of $\alpha$ is presented in Fig.~\eqref{fig:fig3}B. We should mention that this interpretation is valid in steady state. Under dynamical conditions, the parameters of the division propensity are not necessarily related directly to observable statistics. We include the simplification in \eqref{eq:sigmo3} just for stablish intuitive explanations of the role of each parameter on division regulation.

 Next, we perform inference by maximizing the log-likelihood function by optimization, this time, over the parameters $\alpha$, $CV_{\Delta_d|s_b}$ and $\langle \Delta_d\rangle$.~This optimization is performed estimating the probability of division with a similar method as used for obtaining \eqref{eq:prob} but with the propensity function \eqref{eq:sigmo2sb}.

\begin{figure*}[!ht]
\centering
\includegraphics[width=1\linewidth]{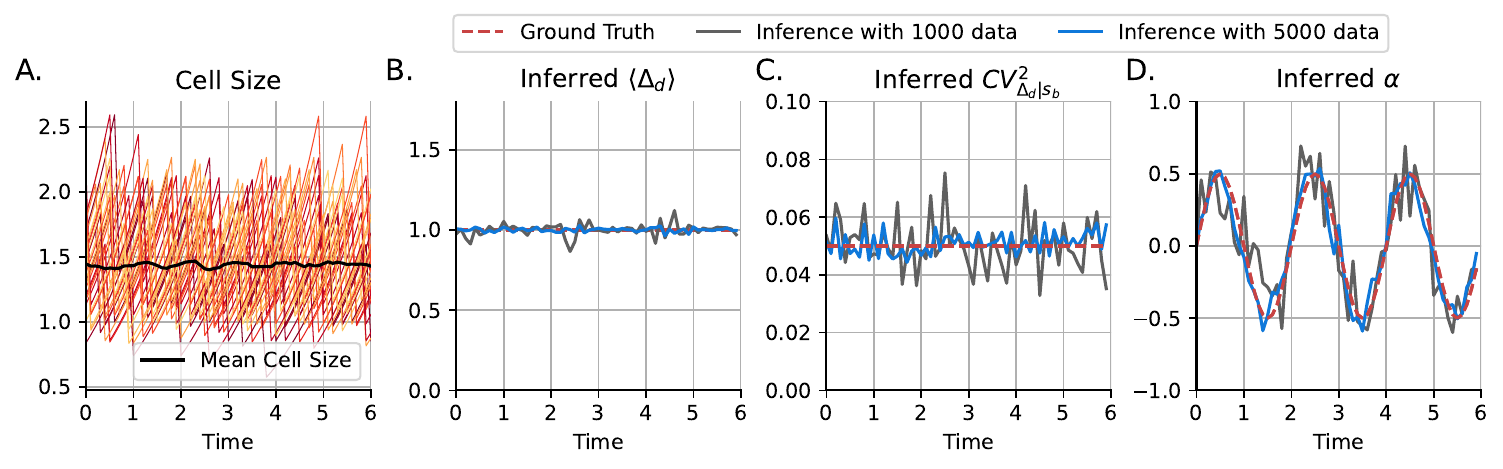}
\caption{{\small \textbf{Inference of a dynamic division strategy ($\alpha$) and comparison between simulated samples of 1000 and 5000 cells.} We simulate two datasets of cells with a dynamical change in the division strategy.~Explicitly, the slope $\alpha$ between $\langle \Delta_d|s_b\rangle$ and $s_b$ changes over time $t$ as $\alpha=0.5\sin(2\pi t)$.~The results of inference of the dataset of 1000 cells is represented in gray.~The inference for the data set of 5000 cells is shown in blue.~(\textbf{A}) Examples of cell size trajectories (different shades of red, and the mean cell size (black)), (\textbf{B}) Inferred $\langle \Delta_d\rangle$, (\textbf{C}) Inferred $CV^2_{\Delta_d|s_b}$ and (\textbf{D}) Inferred $\alpha$.~Other parameters used for the simulation are: $\langle \Delta_d\rangle=\langle s_b\rangle=1$, $CV^2_{\Delta_d|s_b}=0.05$.~}}
\label{fig:fig4}
\end{figure*}

\begin{figure*}[!ht]
\centering
\includegraphics[width=1\linewidth]{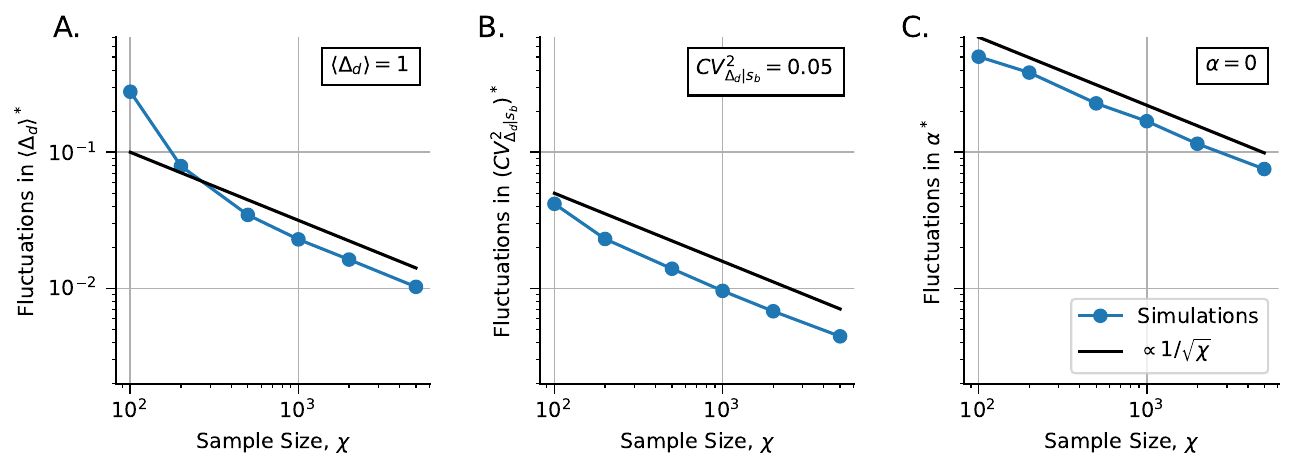}
\caption{{\small \textbf{Fluctuations in the inferred parameters as a function of the sample size used for inference.} We measure the standard deviation of the inferred parameters using 5000 simulated samples for varying sample sizes.~(\textbf{A}) Fluctuations of the inferred $\langle\Delta_d\rangle$.~(\textbf{B}) Fluctuations of the inferred $CV^2_{\Delta_d|s_b}$.~(\textbf{C}) Fluctuations of the inferred $\alpha$.~Other parameters used for the simulation are:  $\langle\Delta_d\rangle=1$, $CV^2_{\Delta_d|s_b}=0.05$ and $\alpha=0$. Black line represents a function that decays as the inverse of the square root of the sample size}}
\label{fig:fig5}
\end{figure*}

\subsection{Performance of the inference method over a dataset with
dynamical division strategy}

To evaluate the performance of our inference method under a time-varying division strategy, we simulated cell populations with a dynamically changing slope parameter $\alpha$, specifically modeled as:
\begin{equation}
    \alpha=0.5\sin(2\pi t).
\end{equation}
We generated two datasets of 1000 and 5000 cells and performed parameter inference on each.
The results are shown in Fig.~\ref{fig:fig4}.~In Fig.~\ref{fig:fig4}A, we present some single-cell size trajectories along with the mean cell size over time.~Inference from 1000 cells (gray curves) captures the general trend of the time-varying division strategy, but exhibits large fluctuations, particularly in the inferred slope parameter $\alpha$ (Fig.~\ref{fig:fig4}D).~Among the inferred parameters, $\langle \Delta_d\rangle$ (Fig.~\ref{fig:fig4}B) shows the least quantitative error, followed by  $CV^2_{\Delta_d|s_b}$  (Fig.~\ref{fig:fig4}C).
In contrast, inference using 5000 cells (blue curves) produces estimates that are more closely aligned with the ground truth (red dashed lines), indicating enhanced precision in all inferred statistics.~These results suggest that larger sample sizes are essential for resolving fine-grained temporal dynamics in the division parameters.

To assess and visualize the reliability of the inferred statistics, we plot the fluctuations in the inferred parameters as a function of the data set size.
For each sample size, we compute the standard deviation of the inferred parameters to quantify their variability (Fig.~\ref{fig:fig5}).~As expected, fluctuations in all three parameters decreased consistently with larger sample sizes with an asymptotical rate inverse of the square root of sample size (black line in Fig.~\ref{fig:fig5}).

\section{Discussion}

In this contribution, we present a simple method for inferring cell division parameters based on maximum likelihood estimation (MLE) methods.~Our framework can estimate three basic parameters: The division scale, represented by either the parameter $\Delta_0$ or the general parameter $\langle \Delta_d\rangle$; the cell size variability represented by either the shape parameter $k$ or by the general parameter $CV^2_{\Delta_d|s_b}$; and the division strategy $\alpha$, which represents the relevance of cell cycle parameters (cell cycle duration, cell size or added size during cell cycle).~The dataset used by our algorithm consists of single-cell size measurements, typically obtained through microscopy-based tracking. Our approach represents a significant advancement over existing methods, particularly in analyzing cellular responses under non–steady-state conditions; an area where current inference algorithms still face major challenges.

The division process is modeled as a stochastic jump with a sigmoidal-shaped division propensity. Although this functional form lacks direct biological interpretation, it provides a tractable framework that yields accurate approximations for the MLE process. The resulting formulas show good numerical stability and include parameters that can be easily related to observable variables such as mean added size and the noise in added size at division. In particular, the proposed division propensity connects naturally to the classical paradigms of sizer, adder, and timer control.

We also examined the robustness of this framework under dynamic changes in parameters. This robustness is important for understanding microbial physiology in fluctuating natural or industrial environments. Previous research has largely focused on steady-state conditions, but real-world nutrient availability is rarely constant. Our results demonstrate that the model can capture 
how division strategies might dynamically shift in response to environmental perturbations, such as resource depletion.
This capability provides possibilities for investigating the new experiments of cell size regulation in fluctuating environments and quantifying how cells adapt to them \cite{miotto2024size,mu2020mass}.~For instance, the model could be used to quantify how a cell transitions from an "adder" to a "sizer-adder" or "timer-adder" strategy as nutrient resources get depleted, providing quantitative insights into the molecular mechanisms governing these adaptive responses\cite{bakshi2021tracking}.

Another important property of our model, is that it can be used to compare cell size regulation across different experimental conditions or genetic backgrounds, even when the complete molecular mechanisms of division are unknown.~This is particularly relevant given the diverse range of growth and division strategies observed across various microorganisms and even human cell lines, which are influenced by factors like nutrient availability, genetic background, metabolic activity, and temperature \cite{panlilio2021threshold,miller2023fission,nguyen2021environmental,knapp2022effects}.~Our framework offers a unified quantitative language to describe these variations.

We performed a systematic investigation of inference accuracy as a function of sample size which can be particularly valuable for experimentalists, guiding the design of future single-cell microscopy studies to ensure sufficient data for reliable parameter estimation.~Future research will focus on applying this MLE framework to diverse experimental single-cell datasets from various microbial species under different dynamic environmental conditions (e.g., varying nutrient limitations, antibiotic exposure, pH shifts)~\cite{cylke2022effects,chure2023optimal}.~This empirical validation will be important to further refine the model and explore the biological significance of the inferred parameters (scale, shape, and division strategy).~Additionally, extending the model to account for other sources of noise, such as measurement error or intrinsic stochasticity in growth, could further enhance its accuracy.~Finally, we plan to explore the predictive power of the inferred single-cell division parameters for understanding population-level dynamics and fitness in competitive environments~\cite{nieto2022cell}.

\begin{credits}
\subsubsection{\discintname}
This work is supported by NIH-NIGMS via grant
R35GM148351.~The funders had no role in study design, data collection and analysis, decision to publish, or preparation of the manuscript.
\end{credits}

\section*{Data Avalability}
Scripts implementing the simulation and inference of cell size dynamics are available in our public repository: 
https://doi.org/10.5281/zenodo.17730929

%
%
%
\bibliographystyle{splncs04}
\bibliography{References}

@article{nieto2021continuous,
  title={Continuous rate modeling of bacterial stochastic size dynamics},
  author={Nieto, Cesar and Vargas-Garcia, Cesar and Pedraza, Juan M},
  journal={Physical Review E},
  volume={104},
  number={4},
  pages={044415},
  year={2021},
  publisher={APS}
}

@article{chure2023optimal,
  title={An optimal regulation of fluxes dictates microbial growth in and out of steady state},
  author={Chure, Griffin and Cremer, Jonas},
  journal={Elife},
  volume={12},
  pages={e84878},
  year={2023},
  publisher={eLife Sciences Publications Limited}
}

@article{knapp2022effects,
  title={The effects of temperature on cellular physiology},
  author={Knapp, Benjamin D and Huang, Kerwyn Casey},
  journal={Annual Review of Biophysics},
  volume={51},
  number={1},
  pages={499--526},
  year={2022},
  publisher={Annual Reviews}
}

@article{cylke2022effects,
  title={Effects of antibiotics on bacterial cell morphology and their physiological origins},
  author={Cylke, Callaghan and Si, Fangwei and Banerjee, Shiladitya},
  journal={Biochemical Society Transactions},
  volume={50},
  number={5},
  pages={1269--1279},
  year={2022},
  publisher={Portland Press Ltd.}
}

@article{miller2023fission,
  title={The fission yeast cell size control system integrates pathways measuring cell surface area, volume, and time},
  author={Miller, Kristi E and Vargas-Garcia, Cesar and Singh, Abhyudai and Moseley, James B},
  journal={Current Biology},
  volume={33},
  number={16},
  pages={3312--3324},
  year={2023},
  publisher={Elsevier}
}

@article{nguyen2021environmental,
  title={Environmental fluctuations and their effects on microbial communities, populations and individuals},
  author={Nguyen, Jen and Lara-Guti{\'e}rrez, Juanita and Stocker, Roman},
  journal={FEMS microbiology reviews},
  volume={45},
  number={4},
  pages={fuaa068},
  year={2021},
  publisher={Oxford University Press}
}

@article{panlilio2021threshold,
  title={Threshold accumulation of a constitutive protein explains E. coli cell-division behavior in nutrient upshifts},
  author={Panlilio, Mia and Grilli, Jacopo and Tallarico, Giorgio and Iuliani, Ilaria and Sclavi, Bianca and Cicuta, Pietro and Cosentino Lagomarsino, Marco},
  journal={Proceedings of the National Academy of Sciences},
  volume={118},
  number={18},
  pages={e2016391118},
  year={2021},
  publisher={National Academy of Sciences}
}

@article{mu2020mass,
  title={Mass measurements during lymphocytic leukemia cell polyploidization decouple cell cycle-and cell size-dependent growth},
  author={Mu, Luye and Kang, Joon Ho and Olcum, Selim and Payer, Kristofor R and Calistri, Nicholas L and Kimmerling, Robert J and Manalis, Scott R and Miettinen, Teemu P},
  journal={Proceedings of the National Academy of Sciences},
  volume={117},
  number={27},
  pages={15659--15665},
  year={2020},
  publisher={National Academy of Sciences}
}

@article{nguyen2021distinct,
  title={A distinct growth physiology enhances bacterial growth under rapid nutrient fluctuations},
  author={Nguyen, Jen and Fernandez, Vicente and Pontrelli, Sammy and Sauer, Uwe and Ackermann, Martin and Stocker, Roman},
  journal={Nature Communications},
  volume={12},
  number={1},
  pages={3662},
  year={2021},
  publisher={Nature Publishing Group UK London}
}

@article{kratz2023dynamic,
  title={Dynamic proteome trade-offs regulate bacterial cell size and growth in fluctuating nutrient environments},
  author={Kratz, Josiah C and Banerjee, Shiladitya},
  journal={Communications Biology},
  volume={6},
  number={1},
  pages={486},
  year={2023},
  publisher={Nature Publishing Group UK London}
}

@inproceedings{cylke2024energy,
  title={Energy allocation theory for bacterial growth control in and out of steady state},
  author={Cylke, Arianna and Serbanescu, Diana and Banerjee, Shiladitya},
  booktitle={Proceedings A},
  volume={480},
  number={2300},
  pages={20240219},
  year={2024},
  organization={The Royal Society}
}

@article{bakshi2021tracking,
  title={Tracking bacterial lineages in complex and dynamic environments with applications for growth control and persistence},
  author={Bakshi, Somenath and Leoncini, Emanuele and Baker, Charles and Ca{\~n}as-Duarte, Silvia J and Okumus, Burak and Paulsson, Johan},
  journal={Nature Microbiology},
  volume={6},
  number={6},
  pages={783--791},
  year={2021},
  publisher={Nature Publishing Group UK London}
}

@article{modi2017analysis,
  title={Analysis of noise mechanisms in cell-size control},
  author={Modi, Saurabh and Vargas-Garcia, Cesar Augusto and Ghusinga, Khem Raj and Singh, Abhyudai},
  journal={Biophysical Journal},
  volume={112},
  number={11},
  pages={2408--2418},
  year={2017},
  publisher={Elsevier}
}

@article{miotto2024size,
  title={A size-dependent division strategy accounts for leukemia cell size heterogeneity},
  author={Miotto, Mattia and Scalise, Simone and Leonetti, Marco and Ruocco, Giancarlo and Peruzzi, Giovanna and Gosti, Giorgio},
  journal={Communications Physics},
  volume={7},
  number={1},
  pages={248},
  year={2024},
  publisher={Nature Publishing Group UK London}
}

@inproceedings{nieto2022cell,
  title={Cell size control shapes fluctuations in colony population},
  author={Nieto, C{\'e}sar and Vargas-Garc{\'\i}a, C{\'e}sar and Pedraza, Juan Manuel and Singh, Abhyudai},
  booktitle={2022 IEEE 61st Conference on Decision and Control (CDC)},
  pages={3219--3224},
  year={2022},
  organization={IEEE}
}

@article{si2019mechanistic,
  title={Mechanistic origin of cell-size control and homeostasis in bacteria},
  author={Si, Fangwei and Le Treut, Guillaume and Sauls, John T and Vadia, Stephen and Levin, Petra Anne and Jun, Suckjoon},
  journal={Current Biology},
  volume={29},
  number={11},
  pages={1760--1770},
  year={2019},
  publisher={Elsevier}
}

@article{jones2023first,
  title={First-passage-time statistics of growing microbial populations carry an imprint of initial conditions},
  author={Jones, Eric W and Derrick, Joshua and Nisbet, Roger M and Ludington, William B and Sivak, David A},
  journal={Scientific Reports},
  volume={13},
  number={1},
  pages={21340},
  year={2023},
  publisher={Nature Publishing Group UK London}
}

@article{cadart2018size,
  title={Size control in mammalian cells involves modulation of both growth rate and cell cycle duration},
  author={Cadart, Clotilde and Monnier, Sylvain and Grilli, Jacopo and S{\'a}ez, Pablo J and Srivastava, Nishit and Attia, Rafaele and Terriac, Emmanuel and Baum, Buzz and Cosentino-Lagomarsino, Marco and Piel, Matthieu},
  journal={Nature Communications},
  volume={9},
  number={1},
  pages={3275},
  year={2018},
  publisher={Nature Publishing Group UK London}
}

@article{lin2018homeostasis,
  title={Homeostasis of protein and {mRNA} concentrations in growing cells},
  author={Lin, Jie and Amir, Ariel},
  journal={Nature Communications},
  volume={9},
  number={1},
  pages={4496},
  year={2018},
  publisher={Nature Publishing Group UK London}
}

@article{taheri2015cell,
  title={Cell-size control and homeostasis in bacteria},
  author={Taheri-Araghi, Sattar and Bradde, Serena and Sauls, John T and Hill, Norbert S and Levin, Petra Anne and Paulsson, Johan and Vergassola, Massimo and Jun, Suckjoon},
  journal={Current Biology},
  volume={25},
  number={3},
  pages={385--391},
  year={2015},
  publisher={Elsevier}
}

@article{singh2010stochastic,
  title={Stochastic hybrid systems for studying biochemical processes},
  author={Singh, Abhyudai and Hespanha, Joao P},
  journal={Philosophical Transactions of the Royal Society A: Mathematical, Physical and Engineering Sciences},
  volume={368},
  number={1930},
  pages={4995--5011},
  year={2010},
  publisher={The Royal Society Publishing}
}

@article{nieto2025generalized,
  title={A Generalized Adder for Cell Size Homeostasis: Effects on Stochastic Clonal Proliferation},
  author={Nieto, C{\'e}sar and Vargas-Garc{\'\i}a, C{\'e}sar Augusto and Singh, Abhyudai},
  journal={Biophysical Journal},
  year={2025},
  publisher={Elsevier}
}

@article{xia2020pde,
  title={{PDE} models of adder mechanisms in cellular proliferation},
  author={Xia, Mingtao and Greenman, Chris D and Chou, Tom},
  journal={SIAM Journal on Applied Mathematics},
  volume={80},
  number={3},
  pages={1307--1335},
  year={2020},
  publisher={SIAM}
}

@article{jia2021cell,
  title={Cell size distribution of lineage data: analytic results and parameter inference},
  author={Jia, Chen and Singh, Abhyudai and Grima, Ramon},
  journal={iScience},
  volume={24},
  number={3},
  year={2021},
  publisher={Elsevier}
}

@article{liu2023cell,
  title={A cell-based model for size control in the multiple fission alga \TEXTIT{Chlamydomonas reinhardtii}},
  author={Liu, Dianyi and Vargas-Garc{\'\i}a, C{\'e}sar Augusto and Singh, Abhyudai and Umen, James},
  journal={Current Biology},
  volume={33},
  number={23},
  pages={5215--5224},
  year={2023},
  publisher={Elsevier}
}

@article{nieto2020unification,
  title={Unification of cell division control strategies through continuous rate models},
  author={Nieto, C{\'e}sar and Arias-Castro, Juan and S{\'a}nchez, Carlos and Vargas-Garc{\'\i}a, C{\'e}sar and Pedraza, Juan Manuel},
  journal={Physical Review E},
  volume={101},
  number={2},
  pages={022401},
  year={2020},
  publisher={APS}
}

@article{nieto2024mechanisms,
  title={Mechanisms of cell size regulation in slow-growing \textit{Escherichia coli} cells: Discriminating models beyond the adder},
  author={Nieto, C{\'e}sar and Vargas-Garc{\'\i}a, C{\'e}sar Augusto and Pedraza, Juan Manuel and Singh, Abhyudai},
  journal={npj Systems Biology and Applications},
  volume={10},
  number={1},
  pages={61},
  year={2024},
  publisher={Nature Publishing Group UK London}
}

@article{ribis2024unique,
  title={Unique growth and morphology properties of {Clade} 5 \textit{Clostridioides difficile} strains revealed by single-cell time-lapse microscopy},
  author={Ribis, John W and Nieto, C{\'e}sar and DiBenedetto, Nicholas V and Mehra, Anchal and Dong, Qiwen and Nagawa, Irene and Meouche, Imane El and Aldridge, Bree B and Dunlop, Mary J and Tamayo, Rita and others},
  journal={bioRxiv},
  pages={2024--02},
  year={2024},
  publisher={Cold Spring Harbor Laboratory}
}

@article{vargas2020modeling,
  title={Modeling homeostasis mechanisms that set the target cell size},
  author={Vargas-Garcia, Cesar A and Bj{\"o}rklund, Mikael and Singh, Abhyudai},
  journal={Scientific Reports},
  volume={10},
  number={1},
  pages={13963},
  year={2020},
  publisher={Nature Publishing Group UK London}
}

@article{jun2018fundamental,
  title={Fundamental principles in bacterial physiology—history, recent progress, and the future with focus on cell size control: a review},
  author={Jun, Suckjoon and Si, Fangwei and Pugatch, Rami and Scott, Matthew},
  journal={Reports on Progress in Physics},
  volume={81},
  number={5},
  pages={056601},
  year={2018},
  publisher={IOP Publishing}
}

@article{amir2014cell,
  title={Cell size regulation in bacteria},
  author={Amir, Ariel},
  journal={Physical review letters},
  volume={112},
  number={20},
  pages={208102},
  year={2014},
  publisher={APS}
}

@article{soltani2016intercellular,
  title={Intercellular variability in protein levels from stochastic expression and noisy cell cycle processes},
  author={Soltani, Mohammad and Vargas-Garcia, Cesar A and Antunes, Duarte and Singh, Abhyudai},
  journal={PLoS computational biology},
  volume={12},
  number={8},
  pages={e1004972},
  year={2016},
  publisher={Public Library of Science San Francisco, CA USA}
}

@article{sanchez2013regulation,
  title={Regulation of noise in gene expression},
  author={Sanchez, Alvaro and Choubey, Sandeep and Kondev, Jane},
  journal={Annual review of biophysics},
  volume={42},
  number={1},
  pages={469--491},
  year={2013},
  publisher={Annual Reviews}
}

@article{shi2021precise,
  title={Precise regulation of the relative rates of surface area and volume synthesis in bacterial cells growing in dynamic environments},
  author={Shi, Handuo and Hu, Yan and Odermatt, Pascal D and Gonzalez, Carlos G and Zhang, Lichao and Elias, Joshua E and Chang, Fred and Huang, Kerwyn Casey},
  journal={Nature communications},
  volume={12},
  number={1},
  pages={1975},
  year={2021},
  publisher={Nature Publishing Group UK London}
}

@inproceedings{rezaee2025inferring,
  title={Inferring cell size control mechanisms through stochastic hybrid modeling},
  author={Rezaee, Sayeh and Nieto, Cesar and Vargas-Garcia, Cesar Augusto and Singh, Abhyudai},
  booktitle={2025 American Control Conference (ACC)},
  pages={1240--1245},
  year={2025},
  organization={IEEE}
}

@inproceedings{nieto2025joint,
  title={Joint Distribution Dynamics of Cell Cycle Variables in Exponentially-Growing Cells with Stochastic Division},
  author={Nieto, C{\'e}sar and Rezaee, Sayeh and Vargas-Garcia, Cesar Augusto and Singh, Abhyudai},
  booktitle={2025 33rd Mediterranean Conference on Control and Automation (MED)},
  pages={108--113},
  year={2025},
  organization={IEEE}
}

\end{document}